\documentclass[manuscript]{aastex61}

\received{\today}
\revised{}
\accepted{}

\submitjournal{ApJ}

\shorttitle{ARs Complex Network}
\shortauthors{Daei, Safari, and Dadashi}

\usepackage{amsmath}
\usepackage{graphicx}
\usepackage{xcolor}
\begin{document}

\title{Complex Network for Solar Active Regions}

\author{Farhad Daei}
\affil{Department of Physics, Faculty of Science, University of Zanjan, P. O. Box 45195-313, Zanjan, Iran}

\author{Hossein Safari}
\affil{Department of Physics, Faculty of Science, University of Zanjan, P. O. Box 45195-313, Zanjan, Iran}

\author{Neda Dadashi}
\affil{Department of Physics, Faculty of Science, University of Zanjan, P. O. Box 45195-313, Zanjan, Iran}

\correspondingauthor{Hossein Safari}
\email{safari@znu.ac.ir}

\begin{abstract}

Here, we developed a complex network of solar active regions
(ARs) to study various local and global properties of
the network. The values of the Hurst exponent ($0.8-0.9$)
were evaluated by both the detrended fluctuation
analysis and the rescaled range analysis
applied on the time series of the AR numbers.
The findings suggest that ARs can be considered as a
system of self-organized criticality. We constructed a growing
network based on locations, occurrence times, and the
lifetimes of 4,227 ARs recorded from 1 January 1999 to
14 April 2017. The behaviour of the clustering
coefficient shows that the ARs network is not a random network.
The logarithmic behaviour of the length scale has the
characteristics of a so-called \textquotedblleft small-world
network\textquotedblright. It is found that the probability
distribution of the node degrees for undirected
networks follows the power-law with exponents
of about 3.7 to 4.2. This indicates the scale-free
nature of the ARs network. The scale-free and small-world
properties of the ARs network confirm that the system of ARs
forms a system of self-organized criticality. Our results show
that the occurrence probability of flares (classified
by GOES class C $> 5$, M, and X flares) in the position of the
ARs network hubs take values greater than that obtained for other
nodes.
\end{abstract}


\section{Introduction}\label{sect1}
Solar active regions (ARs) are the origin of various energetic
phenomena. It is believed that solar flares and coronal mass
ejections are direct results of the changes in the
topology and structure of the ARs' magnetic field
\citep{ref:priest2002,ref:aschwanden}. Because of the important
role of ARs in solar activity, numerous attempts have been
made to study the statistical properties of ARs
\citep{ref:kunzel1959,ref:howard1989,ref:howard2000,
ref:sammis,ref:leka,ref:georgoulis,ref:schrijver,ref:falconer,
ref:mason2010,ref:falconer2011,ref:georgoulis2012,
ref:abramenko2015,ref:Arish2016,ref:barnes2016,ref:raboonic2017}.
\citet{ref:zhang2010} studied both the statistical
({\it e.g.} the frequency distribution of size and magnetic flux)
and physical properties of 1,730 ARs detected from 1996
to 2008. The information (occurrence times and locations on solar
disk) about ARs and sunspots are recorded by the Royal
Observatory of Belgium (ROB), the Solar Influences Data
Analysis Center (SIDC), and the National Oceanic and Atmospheric
Administration (NOAA). In the NOAA catalog, each AR has a unique
identification number. There is important
evidence to suggest that systems of solar events such
as flares, bright points, and so on are in a
self-organized criticality (SOC)
\citep{ref:zhang2001,ref:alipour2015,ref:aschwanden2016}.
\citet{ref:aschwandenbook} provided an extended review
on SOC in solar physics and astrophysics.
\citet{ref:arcangelis2006} studied the universality properties of
both solar flares and earthquakes. They
showed that both phenomena follow the same law for the size
distribution and inter-occurrence times. Also, they interpreted
the temporal power-law dependency of the afterflare
sequence as the Omori law for earthquakes.

The complex network is an approach to study effectively large
complex systems ({\it e.g.} earthquakes, bulk electrical power
systems, computers, the brain, and social systems). This
approach explains significant characteristics of complex systems
using network representations that have their roots in
mathematical studies and are known as the graph theory.
In most studies, a network or its equivalent graph is considered
as a collection of some nodes together with edges while modeling a
network \citep{ref:newman2003,ref:humphries2008,ref:rubinov2010,
ref:lotfi2012,ref:lotfi2013,ref:rezaei2017}. Probability theory
is a useful tool for interpreting the details of the
complexity of the networks. The scale-free and small-world
networks are the two main complex networks
that have been widely used to study large complex
systems ({\it e.g.} \citealt{
ref:watts,ref:amaral2000,ref:barabasi,ref:kim2008,ref:buldyrev2010}).
Small-world networks are basically intersections of the random
and regular networks. In these kinds of networks, the length
scale behaviour is the same as in the random networks,
but with a high average in the clustering coefficient
\citep{ref:watts}. The quantitative measurement of the
small-world property can be computed by different
methods--all comparing the network's parameters with
the parameters of an equivalent random or regular network
\citep{ref:humphries2008,ref:telesford}.
\citet{ref:humphries2008} presented a measure for the small-world
property of a network by comparing both its length scale and
clustering coefficient with an equivalent random network. This
factor is employed in many articles, especially in the
field of neuroscience \citep{ref:bullmore2009}. However,
\citet{ref:telesford} explained that the Humphries
method cannot correctly recognize the small-world networks. They
introduced a new factor, using the length scale of a
random network and the clustering coefficient of a regular
network to evaluate the small-world property of a network.

In this paper, we construct a network for the ARs using their
locations, occurrence times, and lifetimes. We compute the length
scale, clustering coefficient, and degree distribution of nodes
for the network. We describe important properties of the ARs
network in the category of scale-free and small-world networks.

The details are discussed as follows: Section
\ref{sec:data} introduces the data set. In Section \ref{sec:DFA},
the application of both detrended fluctuation and
rescaled range (R/S) analysis on the time
series of ARs is explained. Sections \ref{sec:ARN} and
\ref{sec:NP} describe the construction of ARs networks and their
properties. Sections \ref{sec:results} and
\ref{sec:conclusion} present the results and conclusions,
respectively.

\section{Data Processing}\label{sec:data}
The solar monitor (www.solarmonitor.org) records the solar data observed by several solar
space observatories and missions ({\it e.g.} GOES, GONG, ACE, STEREO, SDO, etc.).

\begin{figure}
\epsscale{1.0} \plotone{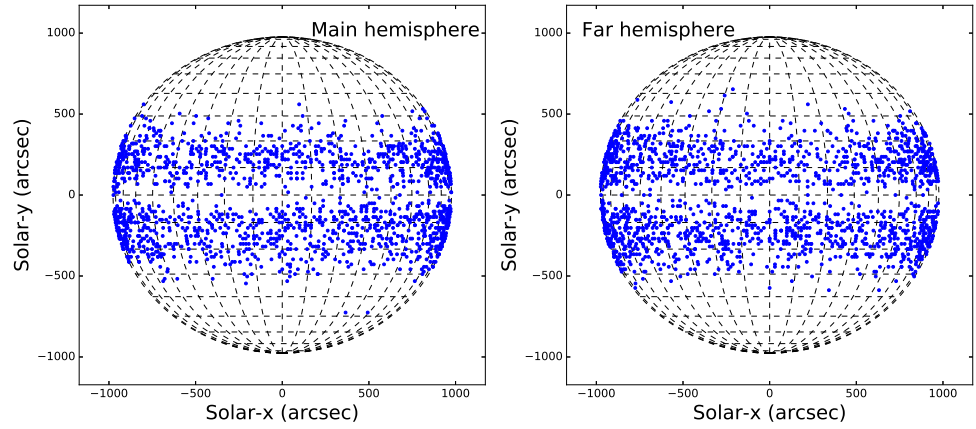} \caption{ Positions of 4,227
ARs (from 1 January 1999 to 14 April 2017) at their first
occurrence times, rotated with respect to the first AR
(NOAA 8419) coordinates. All positions longitudes and
latitudes are restricted to $0^{\circ}-360^{\circ}$ and
$0^{\circ}-180^{\circ}$, respectively, and are mapped to the
surface of a sphere (see text). \label{fig1}}
\end{figure}

SolarMonitor.org provides daily tables for ARs
that occur on the main-side of the
sun. The daily tables include the unique identification
NOAA numbers, sunspot areas, Hale and McIntosh
classifications, produced flares, and the number of spots
\citep{ref:gallagher2002}. The AR positions are
tabulated daily in both the heliographic (in latitude
and longitude) and heliocentric (in arcsec) coordinates. The
needed information of 4,227 ARs during 1 January 1999 to
14 April 2017 for building the ARs network are the NOAA
numbers, rotated positions, occurrence times, and lifetimes. By
tracking the NOAA numbers in daily tables, the lifetimes and
positions of the ARs are extracted. The appendix of this paper
contains a table that includes the NOAA numbers, dates, and
positions in the heliographic coordinates (longitudes and
latitudes) for 4,227 ARs. Using the
diff\_rot code in the SunPy software \citep{ref:sunpy}, the
locations of the ARs are rotated with respect to the first
occurrence time of the reference AR (NOAA 8419). For more
simplicity, we used the location of the ARs at their
first occurrence time during their lifetimes (on the
main-side). In Figure \ref{fig1}, the rotated positions
at the first occurrence time of 4,227 ARs in
the front and far hemispheres are presented.
The rotated positions of ARs (longitudes ($\varphi$) and
latitudes ($\theta$)) are restricted to the range of $0^{\circ
}-360^{\circ }$ and $0^{\circ}-180^{\circ }$, respectively.

Using the GOES flare catalog (
\url{ftp://ftp.ngdc.noaa.gov/STP/space-weather/solar-data/
solar-features/solar-flares/x-rays/goes/xrs/}), we identify ARs
including energetic flares (C$>$5, M, and X class) during 1
January 1999 to 14 April 2017.

\section{Hurst exponent}\label{sec:DFA}
\begin{figure}
\epsscale{.80} \plotone{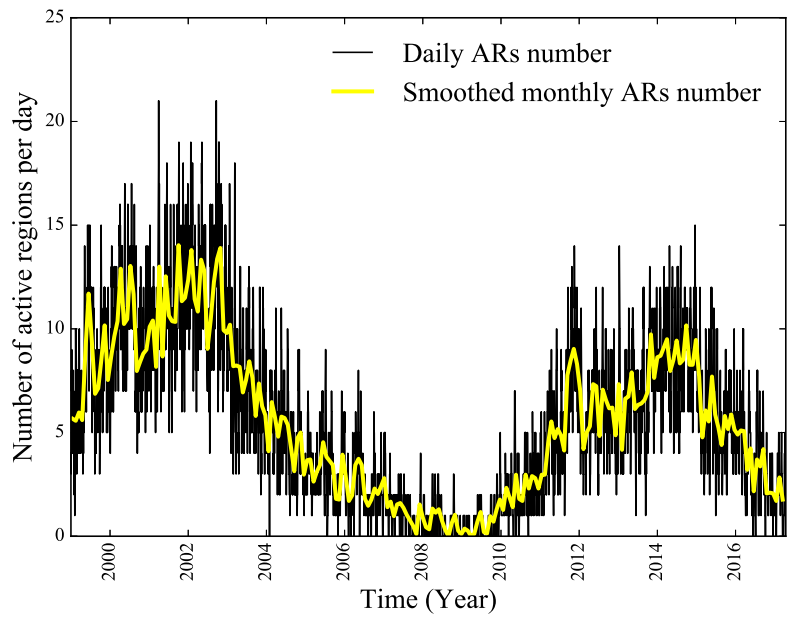} \caption{ Time series of ARs
daily (black line) and smoothed monthly (yellow line) from 1
January 1999 to 14 April 2017 on the solar main-side,
which includes 4,227 ARs. \label{fig2}}
\end{figure}

The Hurst exponent is a key parameter to measure the auto
correlation (self-dependency) of the time series. Both the
detrended fluctuation analysis (DFA)
\citep{ref:mandelbrot1975,ref:peng1994,ref:weron2002,ref:alipour2015}
and R/S analysis \citep{ref:buldyrev1995,ref:hscode} are useful
tools to study the self-affinity (temporal dependency) in the
time series of ARs (Figure \ref{fig2}). In order to compute the
Hurst exponent, the time series of ARs x(t) with length T is
divided into adjacent sub-time series of length P while T=mP.
Each sub-time series is labeled by $x_{i}^{j}(t_{i})~ (i=1,...,P,
j=1,...,m)$. For each sub-time series, the rescaled range (R/S)
is computed by
\begin{eqnarray}
(R/S)_{j} = \frac{\max(y_{1}^{j},...,y_{P}^{j})-\min(y_{1}^{j},...,y_{P}^{j})}{\sqrt{\frac{1}{P-1}\sum_{i=1}^{P}(x_{i}^{j}- \bar{x}^{j})}},
\end{eqnarray}
where $\bar{x}^j=\frac{1}{P}\sum_{i=1}^{P}x_{i}^{j}$ and
$y_{k}^{j}=\sum_{i=1}^{k}(x_{i}^{j}-\bar{x}^{j}),k=1,...,P$. The
average of the normalized range for the P sub-time series is
given by $(R/S)_{P}=\frac{1}{m}\sum_{j=1}^{m}(R/S)_{j}$.
\cite{ref:mandelbrot1975} and \cite{ref:weron2002} have shown
that the R/S analysis shows asymptotic behaviour as $(R/S)_{P}
\propto P^H$, in which H is the Hurst exponent, which can be
obtained from a linear fit in the log-log scale of $(R/S)_{P}$
versus P.

In the DFA for each cumulative time series
$\sum_{i=1}^{k} (x_{i}^{j}-\bar{x}^{j})$, a straight line
$\alpha_{j}t_{i}+\beta_{j}$ is fitted. The mean value of the mean
square fluctuation for all sub-time series is computed using
\begin{equation}
F(P)=\frac{1}{m}\sum_{j=1}^{m}\left(\frac{1}{P}\sum_{i=1}^{P}((x_{i}^{j}-\bar{x}^{j}) -
\alpha_{j}t_{i}-\beta_{j})^2\right)^{1/2}.
\end{equation}
The Hurst exponent (H) can be obtained by a linear fit
(in a log-log scale) for F(P) versus P.

DFA and R/S analyses are used to extract the value of
the Hurst exponent (H) to investigate the behaviour of the time
series. In the analysis of time series based on the value of the
Hurst exponent (H), if $0<H<0.5$ and $0.5<H<1$, we can say the
time series has long temporal negative and positive correlations,
respectively. If H equals to 0.5, the time series is
uncorrelated (white noise) \citep{ref:buldyrev1995}. By applying
the DFA and R/S on the time series of ARs, the values of the
Hurst exponent are obtained as 0.8 and 0.94, respectively.
These values show that the time series of ARs has a long temporal
dependency and suggest that the system of ARs is one of
SOC \citep{ref:carreras2001,ref:dobson2007,ref:alipour2015}.
SOC is an important way to describe the
complex nature of a high degree of freedom and
non-linearity in many physical and astrophysical phenomena ({\it
e.g.}
\citealp{ref:bak1987,ref:tang1988,ref:wang2013,ref:aschwanden2016})

 In the remainder of this paper, we study the behaviour of the
SOC of the ARs system in the framework of the complex
network.

\section{Active Regions Network}\label{sec:ARN}
\begin{figure}
\epsscale{.80} \plotone{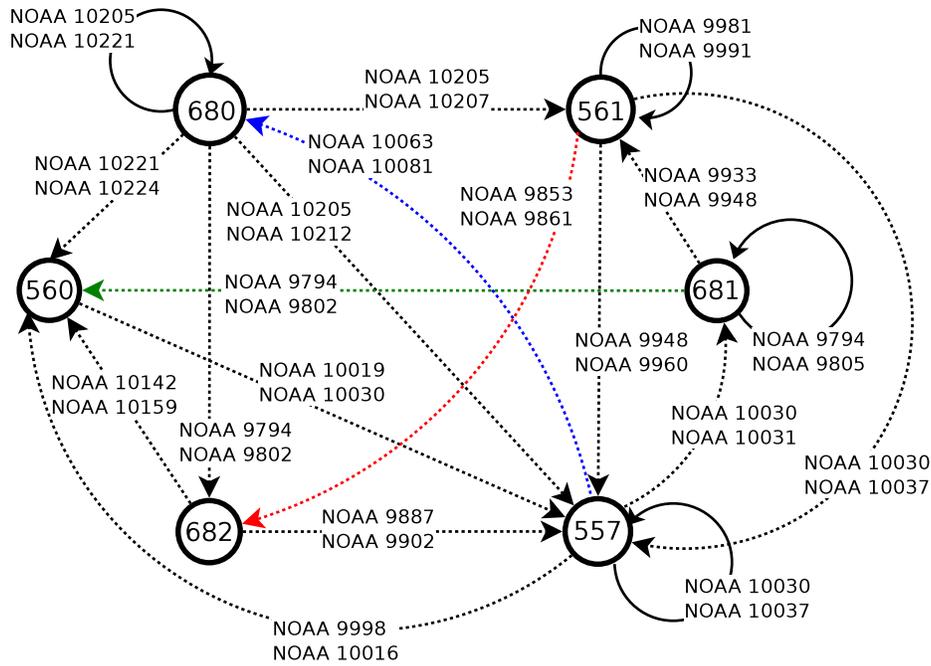} \caption{A small part of
the ARs network comprising six nodes is
represented. Each arrow represents the edge between the two
nodes. For example, the AR (NOAA 10081) emerges in node $n_{680}$
during the lifetime of AR (NOAA 10063) presented in node
$n_{557}$, so that a directed edge is drawn from $n_{557}$ to
$n_{680}$. Each self-loop shows the occurrence of two
coinciding ARs in the same node. For example,
the AR (NOAA 10205) appearing at the same node
of AR (NOAA 10221) formed the self-loop. \label{fig3}}
\end{figure}

The information of 4,227 recorded ARs is used
to construct a complex network concerning the following steps:
\begin{itemize}
\item[--] The solar spherical surface is divided into $N \times N$ cells with equal areas
considering the spherical coordinates ($\theta,\varphi$) as
$A_{\imath,\jmath}=\frac{4\pi}{N^{2}}R_{\odot}^{2}\; ,\; \imath , \jmath= 1 .. N$,
where the parameter $R_{\odot}$ is the solar radius. The angles $\theta$ and $\varphi$ for
each equal area cell are given by:
\begin{eqnarray}
&&\varphi_{\imath + 1 } = \varphi_{\imath } + \frac{2\pi}{N}\;,\; \varphi_{1} = 0,\nonumber \\
&&\cos(\theta_{\jmath + 1 } ) = \cos(\theta_{\jmath }) - \frac{1}{N}\; , \; \theta_{1} = 0.
\end{eqnarray}

\item[--] The ARs are assigned to cells according to their
locations at their first occurrence times rotated with respect to
the first occurrence time of the reference AR (NOAA 8419) (Figure
\ref{fig1}). To avoid more complexity in our analysis, the
variations on the latitudes and longitudes of ARs during their
lifetimes are ignored. Cells with no ARs are removed from our
network analysis. However, the remaining ones are used as the
nodes of the network, indicated by $n_{\ell}\: (\ell = 1
.. L)$.

\item[--] An edge connects the node $n_{i}$ to $n_{j}$ if an AR appears at
node (cell) $n_{j}$ during the lifetime of another AR at node
$n_{i}$. A loop (connecting node $n_{i}$ to itself) is formed
when an AR appears during the lifetime of another one within the
same cell. In Figure \ref{fig3}, a small part of the ARs network
with six nodes, 15 edges, and four loops is
shown. This graph presents a sample of our network, which
contains the information of 33 ARs. The AR NOAA 10081 is formed
during the lifetime of AR NOAA 10063; therefore, in our network,
the node number 680 is connected to the node
that is labeled by 557. Also, the AR NOAA 9805 appeared
in the lifetime of AR NOAA 9794 at the same node (681); so, a loop
is created.

\item[--] We organize a directed and weighted network as
an $L \times L$ adjacency matrix ($A^{\prime }$). In
order to analyze the properties of the ARs network,
$A^{\prime }$ is converted to a symmetric matrix with all
diagonal elements equal to zero, which is representative of the
undirected, unweighted, and self-loop-free graph.

\end{itemize}

\section{Network Parameters}\label{sec:NP}

A graph that consists of nodes and edges is a mathematical
representation of a network. In general, a graph can be
categorized into a directed or an undirected
and a weighted or an unweighted graph
depending on its edges. A directed network or its equivalent
graph is defined by a set of nodes including directed connections
(edges). A graph with bi-directional edges is called an
undirected graph. A graph with different real numbers assigned to
its edges is named a weighted network. The unweighted
ones are those for which all the weights are set to be equal to
one or the edges have no number assigned to them. In the complex
network approach, using the adjacency matrix made for a graph, the
topological properties in both the local and global scales were
studied \citep{ref:steen}. In the present section, we briefly
review some network parameters.

In order to characterize a graph, one can use an adjacency matrix
($A^{\prime}$), including information about edges and nodes.
The adjacency matrix is an $L \times L$ matrix
in which the element $a_{i,j}^{\prime}$ is the number of edges
({\it i.e.} weight of edge) connecting node $n_{i}$ to node
$n_{j}$. In this way, the sum of $i^{\rm th}$ row elements
represents the outbound edges from the node $n_{i}$
(outbound degree) and the sum of the $i^{\rm th}$ column elements
represents the number of inbound edges to the node
(inbound degree). The diagonal elements represent
self-loop(s) of nodes ({\it i.e.} a node connecting to itself by
an edge). For a simple undirected and unweighted graph,
the adjacency matrix is a symmetric matrix with all
elements equal to zero or one. Undirected and
unweighted adjacency matrices are used to determine the
characteristics of the ARs network, including the
average of local clustering coefficients, mean of shortest path
length, diameter, and the nodes' degree of probability
distribution function.

To convert the network to an unweighted, undirected, and
loop-free network, the diagonal elements of the adjacency matrix
are set to zero and every non-zero element is replaced by
one. Then, the new adjacency matrix (A) is
symmetrized.

The local clustering coefficients ($c_{\imath}$) and
clustering coefficient (C) are computed using the adjacency matrix A:
\begin{eqnarray}\label{eq:C}
&&c_{i} = \frac{\sum_{j,\ell \neq i} a_{i,j} a_{j,\ell} a_{\ell,i} }{k_{i}(k_{i}-1)},\\
&&C = \frac{1}{L} \sum_{i=1}^{L} c_{i},
\end{eqnarray}
where $k_{i}$ and L are the number of edges connected to the $i^{th}$ node and the total number
of nodes, respectively.

The average of the shortest path between all pairs of nodes is another interesting
parameter of the network, expressed as:
\begin{align}\label{eq:ls}
ls = \frac{1}{L(L-1)} \sum_{i,j= 1, i \neq j}^{L} d_{i,j},
\end{align}
where $d_{i,j}$ is the shortest path length between i and j
nodes.

The parameter $d_{i,j}$ is computed using Floyd--Warshall
algorithm \citep{ref:floyd1962}, which is an efficient method to
find the length of the shortest path between all nodes
in a graph. The largest value of the shortest paths is called
the diameter of a graph. Another interesting property of
a network is the probability distribution function (PDF) of
the nodes' degree, which determines the probability
of finding a node with a certain degree.\\
In this study, we refer to some properties of such well-known networks as regular, random,
small-world, and scale-free networks.

In regular graphs with all nodes having the same degree,
typically the clustering coefficient and shortest path
length possess large values. A network is called random if all
pairs of nodes are connected with the same probability
\citep{ref:ermodel1960}. A random graph, in comparison
with a regular one, has smaller values for
both the clustering coefficient and the shortest path length. For
a random network, the average values of the shortest path length
($ls_{{\rm rand}}$) and clustering coefficient ($C_{{\rm rand}}$)
can be obtained by the following equations:
\begin{eqnarray}\label{eq:crand}
&&ls_{{\rm rand}} = \frac{\ln(L_{{\rm rand}}) - \gamma}{\ln{<k>}}+\frac{1}{2},\\
&&C_{{\rm rand}} \simeq \frac{E_{{\rm rand}}}{L^{2}_{{\rm rand}}},
\end{eqnarray}
where $E_{{\rm rand}}$, $L_{{\rm rand}}$, $<k>$, and
$\gamma=0.5772$ are the number of nodes, number of
edges, average degree of nodes, and Euler constant,
respectively \citep{ref:fronczak2004}. One can build a random
graph using a model proposed by \citet{ref:ermodel1960}. The
intersection of the random and regular graphs is called
a small-world graph \citep{ref:watts}. A small-world
graph has a small average shortest path. However, the clustering
coefficient takes the larger value. We can find enough examples
of such graphs in nature--for instance, in the
brainstem reticular network \citep{ref:humphries2005}, human
protein network \citep{ref:stelzl2005957}, and power grid network
\citep{ref:mei2011}. \citet{ref:bullmore2009} reviewed the
small-world networks in neurosciences. In some small-world
networks ({\it e.g.} the brainstem reticular network),
the PDF of the nodes' degree (P(k)) follows a
power-law, $P(k) \propto k^{-\gamma}$. This kind of
network is representative of scale-free events.

\section{Results}\label{sec:results}
In order to construct a complex network for the solar ARs, the
locations and dates of the first occurrence times, and the
lifetimes of 4,227 ARs were used (Section
\ref{sec:ARN}). The nodes of network was created by dividing the
solar surface into equal area cells. The adjacency matrix
was prepared for both the directed and undirected ARs
networks. The weighted adjacency matrix was determined using
directed connections (edges) between the pair of ARs.
Loops, the connections between two successive ARs occurring at
the same cell (node), were considered as the diagonal elements of
the adjacency matrix ($a_{i, i}^{\prime}$). The empty cells of
the directed and weighted adjacency matrix were removed from our
network analysis. The effect of the network size (the number of
nodes, L) on the shortest path length, clustering coefficient,
and probability distribution of the nodes' degree are
studied. An equivalent random network was constructed
corresponding to the ARs network with the same size and edges.

\begin{figure}
\epsscale{.80} \plotone{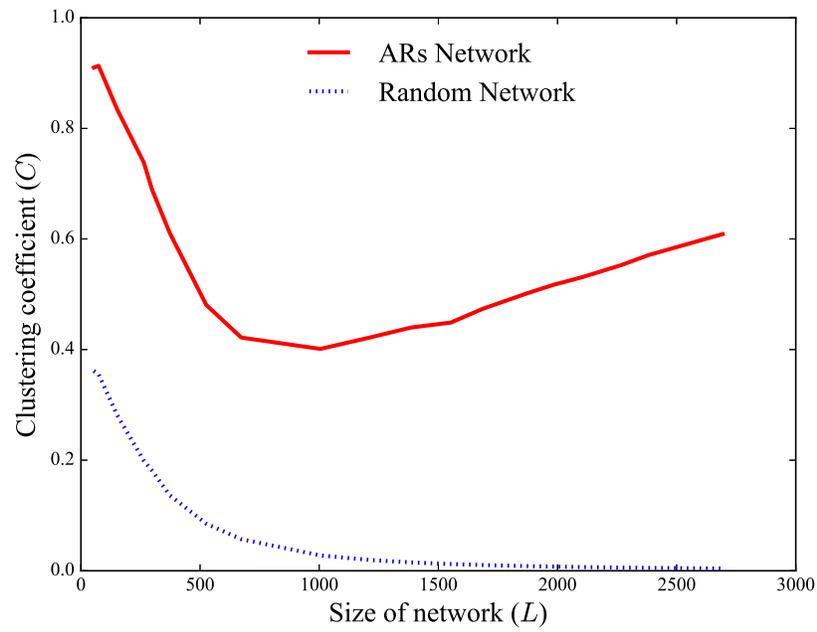} \caption{Clustering coefficient
of ARs networks and their equivalent random networks (with the
same number of nodes and edges) are plotted. The size of
the network ranges from 50 to 2,700 (nodes).
\label{fig4}}
\end{figure}

The clustering coefficients of the ARs networks and their
equivalent random networks (with the same number of nodes and
edges) are shown in Figure \ref{fig4}. The size of the networks
varies from 54 to 2,693 (nodes). The clustering coefficient
for the ARs and random networks are computed by Eqs. (\ref{eq:C})
and (\ref{eq:crand}), respectively. It is found that the
clustering coefficient of the ARs network is noticeably greater
than the equivalent random network by about a factor of two or
more.

\begin{figure}
\epsscale{.80} \plotone{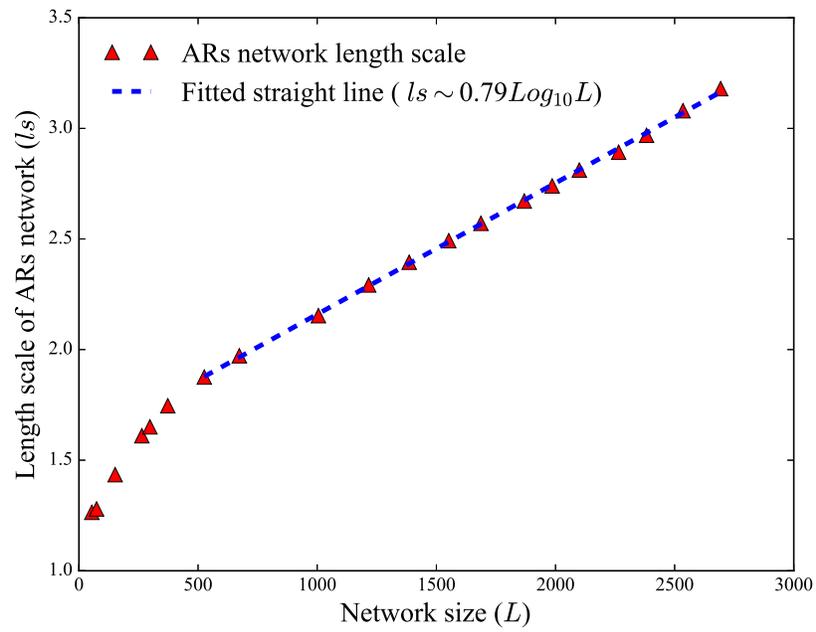} \caption{Length scale of the
ARs network (triangle) versus the network size (L) and a fitted
straight line as $ls \sim 0.79\: \log_{10}L$ are plotted.
\label{fig5}}
\end{figure}

In Figure \ref{fig5}, the length scale of the ARs network versus
the size (logarithm) of the network is plotted. For small size of the ARs
networks ($L < 500$) most of the nodes are connected together and the networks
tend to the complete graphs. For the networks with size 500-2,700 a linear fitting
was applied and the slope of the fit was obtained ($ls = 0.79\:
\log_{10}L$). According to the small values of the length scale
and its logarithmic dependency on the size, we conclude
that the ARs network is classified as a small-world network.

\begin{figure}
\epsscale{1.1} \plotone{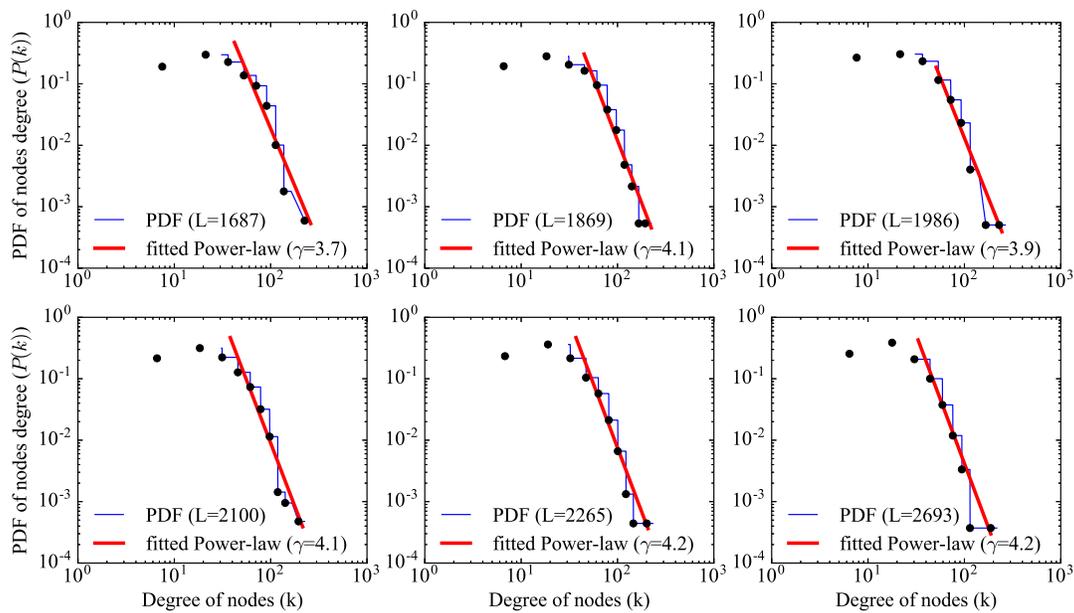} \caption{Probability
distribution function (PDF) for the nodes' degrees
(circles) for undirected ARs network are plotted in a
log-log scale. The exponent for the fitted truncated
power-laws (dashed line) is obtained in the range of
3.7-4.2. \label{fig6}}
\end{figure}

Figure \ref{fig6} presents the PDF of the nodes' degree
for the undirected networks with the size varying from 1,687 to
2,693. The lower ends of the degree distributions ($k < k_{trunc}$) suffer from some
truncation effects due to incomplete detection of small ARs and far-side ARs,
therefore should not be
considered in the power-law fitting. The truncated power-law ($p(k) \sim
k^{-\gamma}$) fitting was carried out on the probability
distribution of the nodes' degree \citep{ref:Aschwanden2015}. The
values of the power-law exponent range from 3.7 to 4.2.
The power-law nature of the PDF indicates that a few
nodes have high connectivity values. In the context of
the complex network, nodes with a large number of
connections are named hubs of the network \citep{ref:albert2002}.
One may ask: What is the criterion for selecting a node
as a hub? In a network, nodes with degrees higher than a
threshold are considered as hubs. The threshold is defined as the
maximum degree of nodes in the equivalent random network.

\begin{figure}
\epsscale{.80} \plotone{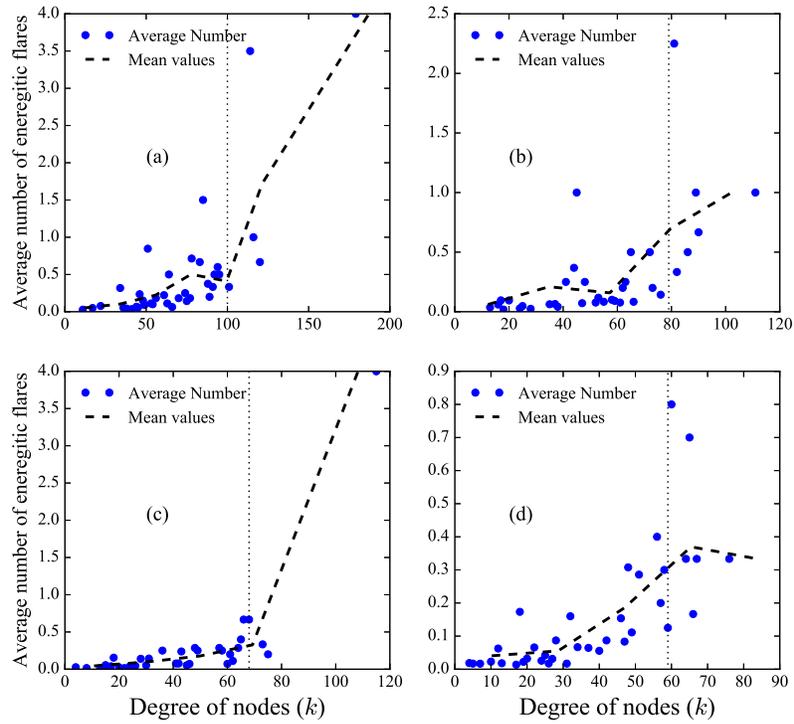} \caption{The average number of
flares (circles) and connected mean values of non-uniform binning
(dashed line) observed within the position of the ARs network
nodes versus the degree of nodes for the network size (L) (a)
1,552, (b) 1,986, (c) 2,382, and (d) 2,693 are plotted.
The nodes with high degrees (hubs) are presented on the right
side of the dotted lines. \label{fig7}}
\end{figure}

\begin{figure}
\epsscale{.80} \plotone{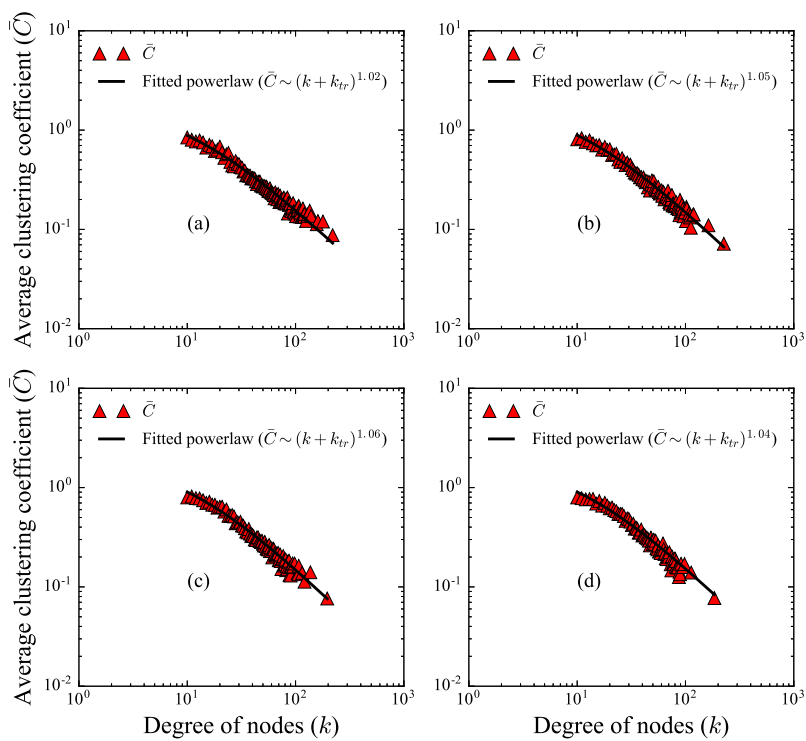} \caption{The average of the
clustering coefficients of nodes with the same degree (triangles)
in the undirected ARs networks and the fitted threshold power-laws ($ C \sim (k+k_{th})^{-\gamma}$ ) are presented for
network sizes (a) 1,552, (b)
1,986, (c) 2,382, and (d) 2,693. The value of the threshold ($k_{th}$) selected
to be 10 and the fitted power-law exponents obtained to be in the range of $1.02 - 1.06$.
\label{fig8}}
\end{figure}

We found that for the ARs network with 1,986 nodes,
there are 53 hubs with degrees larger than 80. On
increasing the network size to 2,693 nodes, 78 hubs
with connections larger than 61 are found. In Figure \ref{fig7},
the relation between the average number of large energetic flares
(C$>5$, M, and X classes) appearing at the position of
the ARs network nodes is presented. The average number of flares
increases with an increase in the degree of
the nodes. The average number of large flares
appearing at the position of the ARs network hubs is
larger than the other nodes by at least a factor of two. In
Figure \ref{fig8}, the dependency of the averaged local
clustering coefficient ($\bar{C}$) for the nodes with the same
degree is presented.The threshold
power-law($\bar{C}\sim(k+k_{\rm th})^{-\gamma}$) is fitted to
the average local clustering coefficient in which $k_{\rm th}$ is
the threshold value. \citet{ref:Aschwanden2015} modeled the
threshold power-law for size distributions of some natural
phenomena ({\it e.g.,} solar and stellar flares, etc.). They show that
in most size distributions of detected data, the threshold
power-laws are significantly well fitted. We found that
by increasing the size of the network, the
power-law exponent of the average clustering
coefficient slightly exceeds the value one.

\section{Conclusion}\label{sec:conclusion}
Typically, 220 ARs can be observed on the solar disk each year.
Within some of the ARs, large-scale magnetic phenomena such as
flares and coronal mass ejections are stochastically emerged. The
exact physical mechanism underlying these phenomena
remains unknown. The results of applying both the
detrended fluctuation and R/S analysis with the Hurst
exponent ($0.8-0.9$) suggest that the ARs system is
categorized into SOC, which motivated us to
confirm such a characteristic by using the
complex network approach. In the present work, we designed the ARs
network based on their locations at the first occurrence
times and lifetimes. In our complex network, all ARs
that occurred in the lifetime of a specific AR are
linked with each other. The unweighted adjacency matrix
was used to calculate the length scale and clustering coefficient
of the network. The degree of nodes was computed for both the
directed and undirected ARs networks. The main results
of this study are organized as follows:
\begin{itemize}
\item[--] A comparison between both the values and behaviour of the
clustering coefficient of the ARs network and the equivalent
random network (with the same number of edges and nodes)
indicates that the ARs network is not a random network.

\item[--] The small values of the length scale and its logarithmic
behaviour ($ls = 0.74\: \log_{10}L$) related to the
network size demonstrate that the ARs network is a small-world
network. The obtained results for the large values of the
clustering coefficient confirm that the ARs network can be
classified in a category of the small-world network.

\item[--] The truncated power-law distribution was fitted to the PDF
of the nodes' degree. The power-law nature of
the nodes' degree demonstrates the scale-free
feature of the ARs network. Recent observations
have shown that some of the ARs produce energetic
phenomena ({\it e.g.} flares, CMEs, etc) and some others
do not. Such characteristics of the ARs
network are representative of the prescription of the
heterogeneous networks ({\it e.g.} \citealt{ref:abe2009}).

\item[--] We observed an increase in flare occurrence in each cell corresponding to
the hubs. In other words, the nodes with higher degrees in the ARs
network have a higher likelihood to trigger flares
(Figure \ref{fig7}). This behaviour of the ARs network
raises an important question about the prediction
capability of flares based on the ARs network. More
statistical studies are required to examine the flare prediction
with the ARs' complex network.

\end{itemize}
In the present study, we used the ARs information
appearing on the main-side of the solar
surface. By employing the tracking algorithms on the reconstructed
helioseismic maps \citep{ref:lindsey2000,ref:lindsey2013}, the
returning far-side ARs can be identified. By increasing
the lifetimes of some ARs from two weeks to one month or more, the
number of connections (edges and loops) for some nodes of the ARs
network may be increased. Additionally, some of
the nodes may be empty and, therefore, will
be removed from the analysis. Expectedly, by decreasing
the number of nodes in the presence of the ARs with the
large lifetimes, the diameter and the mean path length of the
network may also decrease while the clustering
coefficient may increase. More quantitative studies are
required to address the effects of the far-side ARs information
on the properties of the network.

In the next step, we attempt to include more
characteristics of ARs ({\it e.g.} magnetic class, Hall class,
McIntosh class, sunspots) and study the effects of changing solar
latitudinal and longitudinal displacements of the ARs during
their lifetimes in order to investigate the prediction
capability for energetic solar events.

\section*{Acknowledgement}

The authors thank the unknown referee for his/her very helpful
comments and suggestions. Data supplied courtesy of
SolarMonitor.org.

\clearpage

\clearpage

\clearpage

\clearpage

\clearpage

\clearpage

\clearpage

\clearpage

\appendix

\begin{deluxetable}{cccccc}
\tablecaption{Solar Active Regions data}
\tablehead{
\colhead{ID} & \colhead{Year} & \colhead{Month} &
\colhead{Day} & \colhead{Lat.} & \colhead{Long.}
}
\startdata
8419&1999&01&01& 28& 91 \\
8419&1999&01&02& 28& 77 \\
8420&1999&01&01& 20& 43 \\
8420&1999&01&02& 20& 40 \\
8420&1999&01&03& 20& 39 \\
8421&1999&01&01& 27& 40\\
\enddata
\tablecomments{Table 1 is published in its entirety in the electronic 
edition of the {\it Astrophysical Journal}. A portion is shown here 
for guidance regarding its form and content.}
\end{deluxetable}

\bibliographystyle{aasjournal}
\bibliography{refs}
\end{document}